\documentclass[useAMS,usenatbib]{mn2e}
\usepackage{graphicx}

\newcommand{\kms}{km~s$^{-1}$}
\newcommand{\ms}{{\rm M_\odot}}

\newcommand{\lbol}{L_{\rm bol}}
\newcommand{\halpha}{H$\alpha$}
\newcommand{\urad}{U_{\rm rad}}
\newcommand{\ub}{U_{\rm B}}

\newcommand{\tauloss}{\tau_{\rm loss}}
\newcommand{\diff}{{\rm d}}
\newcommand{\hii}{H\,{\sc ii}\ }

\newcommand{\AandA}{A$\&$A}

\title[Convective wind in NGC~1569]
{Multifrequency radio continuum observations of NGC~1569:
Evidence for a convective wind}
\author[U. Lisenfeld et al.]
{U. Lisenfeld$^1$\thanks{e-mail: ute@iaa.es},
T.W. Wilding$^2$, G.G. Pooley$^2$,
P. Alexander$^2$ \\
$^1$Instituto de Astrofisica de Andalucia (CSIC),
Aptdo. 3004, 18080 Granada, Spain \\
$^2$Astrophysics,
Cavendish Laboratory, Madingley Road, Cambridge, CB3 0HE }

\begin{document}

\date{Accepted ..... Received ...; in original form ...}

\pagerange{\pageref{firstpage}--\pageref{lastpage}} \pubyear{2002}

\maketitle

\label{firstpage}

\begin{abstract}
We present high-sensitivity radio continuum observations
with the VLA and Ryle Telescope at 1.5, 4.9, 8.4 and
15.4 GHz of the dwarf irregular galaxy NGC~1569. The radio data show an
extended, irregularly-shaped halo with filamentary structure around the
galaxy. The spectral index maps 
reveal an unusually patchy distribution with regions of flat spectral index
extending into the halo.
The data allow us to perform a spatially-resolved
spectral fitting analysis of the continuum emission from which we
derive maps of the  thermal and synchrotron emission.
The thermal
radio emission is concentrated towards the brightest \hii region
west of  the super star clusters A and B whereas the distribution
of the synchrotron emission peaks in a
bar-like structure in the disk extending between the two clusters.
The total flux density of the thermal radio emission allows us to
derive the integrated synchrotron spectrum and we confirm the break in
the spectrum that was found by \citet{isr-deb88}.
We discuss various possibilities that could produce such a break and
conclude that the only mechanism able to fit the radio
data and being consistent
with data at other wavelengths is a convective wind allowing
cosmic ray electrons to escape from the halo.
\end{abstract}


\section{Introduction}

NGC~1569 is a nearby dwarf irregular galaxy (we adopt a distance of
$2.2 \pm 0.6$ Mpc,
\citealt{isr88}),
currently in the aftermath of a
starburst.
The high star-formation rate (SFR)
has produced numerous young star-clusters \citep{hun00}, the most
luminous being the super star clusters (SSC) A and B \citep{abl71}.
The interstellar medium (ISM) reflects
the imprints of the active star formation in different ways.
The atomic gas exhibits an unusually high velocity
dispersion \citep{sti00} and there is a remarkable hole in the
HI distribution, situated close
to SSC A \citep{isr90}.
The \halpha \ emission is particularly unusual with a very filamentary
structure, reminiscent of an explosion \citep{wal91}, and showing a
complex velocity field consistent with the presence of many
bubbles \citep*{tom94}.
From the kinematics of the \halpha \ filaments 
(\citealt{vau74}, \citealt{wal91}, \citealt{hec95}, \citealt*{mar02})
 and from X-ray emission
(\citealt{hec95}, \citealt{mar02}) the presence of an
outflow of gas can be inferred.
A possible trigger for the starburst
is the interaction with a $7 \times 10^6$ $\ms$ HI cloud at
5 kpc from NGC~1569 \citep{sti98}.

Numerous studies at different wavelengths have
shed light on the star formation history.
The analysis of colour-magnitude diagrams shows an increased level of
star formation (SF) starting about 100 Myr ago and stopping 4 -- 10 Myr
ago (\citealt{val96}, \citealt{gre98}).
A drop of the SFR about
4 -- 10 Myr ago has also been inferred by \citet{hec95}
from the paucity of stars more massive
than about 20 -- 25 $\ms$ and from the low excitation temperature of
the ionized gas derived from spectroscopy.
Assuming a Salpeter initial mass function (IMF),
with a slope of 2.35, and lower and upper mass cut-offs
at 0.1 and 120 $\ms$,
\citet{gre98} found a SFR of 0.5 $\ms$ yr$^{-1}$ with little
change during the starburst phase (for an IMF slope of 2.6 the rate
increases to 1 $\ms$  yr$^{-1}$). This SFR is very high compared to
other dwarf galaxies \citep{gre98}.
\citet{hun00}, on the other hand, found that most
stellar clusters identified from HST observations had ages of less than
30 Myr, and suggested on this basis that most of the
star formation took place towards the end of the starburst period.
A similar conclusion was reached by \citet{and03}, who analysed
multi-wavelength HST observations of a large number of star clusters
in NGC~1569 and found from their derived age distribution that a major burst
started about 25 Myr ago.

Although the SFR has declined during the last 4 -- 10  Myr, it is still high,
at least at some localized positions,
as demonstrated by the large number of compact \hii \ regions \citep{wal91}
and the young age of only a few Myr derived for SSC A
\citep{gon97}.
\citet{wal91} derives from the \halpha \ data a present SFR of 0.4
$\ms$  yr$^{-1}$
(assuming an ``extended Miller-Scalo IMF'' with a slope of 2.5 above
1 $\ms$, \citealt{ken83}). The inferred SFR would be reduced by a factor of 2
when taking into account  only  \halpha\ emission from \hii\ regions
(i.e. neglecting diffuse \halpha\ emission) and
thus directly linked to massive star formation.
This shows that the SFR has not decreased by more than a factor of about
3--5 during the last $\sim$ 10 Myr compared to the last 100 Myr.

In this paper we present high-resolution
radio continuum data at 4 wavelengths for NGC~1569.
These data allow us to decompose the radio emission into its two
components: free-free emission which is characterized by
a spectral index\footnote{In this paper we define the spectral index  via
$S(\nu)  \propto \nu^{-\alpha}$,
where $ S(\nu)$  is the flux density at frequency $\nu$.}
 of $\alpha=0.1$ and that originates principally in \hii regions, heated by
stars with masses above about 20 $\ms$, and  synchrotron emission
from diffuse relativistic cosmic ray electrons (CREs).
The synchrotron spectrum
is steeper, with $\alpha$ typically between 0.5 and 1.0, than that
of the free-free emission and therefore dominates the radio spectrum
at low frequencies, typically below 10 GHz.
We assume that the main sites for the acceleration
of cosmic rays (CRs) are supernova remnants (SNRs)
associated with
Type II and Ib supernovae (SNe). Such SNe have progenitors
with masses above 8 $\ms$
(\citealt{ken84}, \citealt{bar92}) with life times
of up to a few times $10^7$ yr.
After their injection into the ISM,
CREs suffer energy losses
mainly due to  synchrotron and inverse Compton emission,
which make their energy and radio spectra steepen in a
characteristic fashion on typical time-scales around $10^7$ yr.

Thus, the fitting of the radio continuum maps allows us to
trace the sites of present SF
(positions with free-free emission)
and probe the time since the CREs
were last accelerated,
 and hence previous epochs of SNe and SF.
High frequency observations, preferably above 10 GHz,
 are necessary for a reliable
separation of the thermal and non-thermal radio emission.
Previous studies with equivalent data sets
were done for the starburst galaxy
NGC~2146 \citep{lis96}, the giant spiral galaxy NGC~1961
\citep{lis98} and the interacting galaxies NGC~4490/4485
\citep*{cle99}.

\section{Observations and data reduction}

\begin{figure*}
\begin{minipage}{178mm}
\centerline{\hskip-0.5cm 1.5~GHz \hskip8.5cm    4.9~GHz}
\hbox{
{\rotatebox{270}{\includegraphics[width=6.5cm]{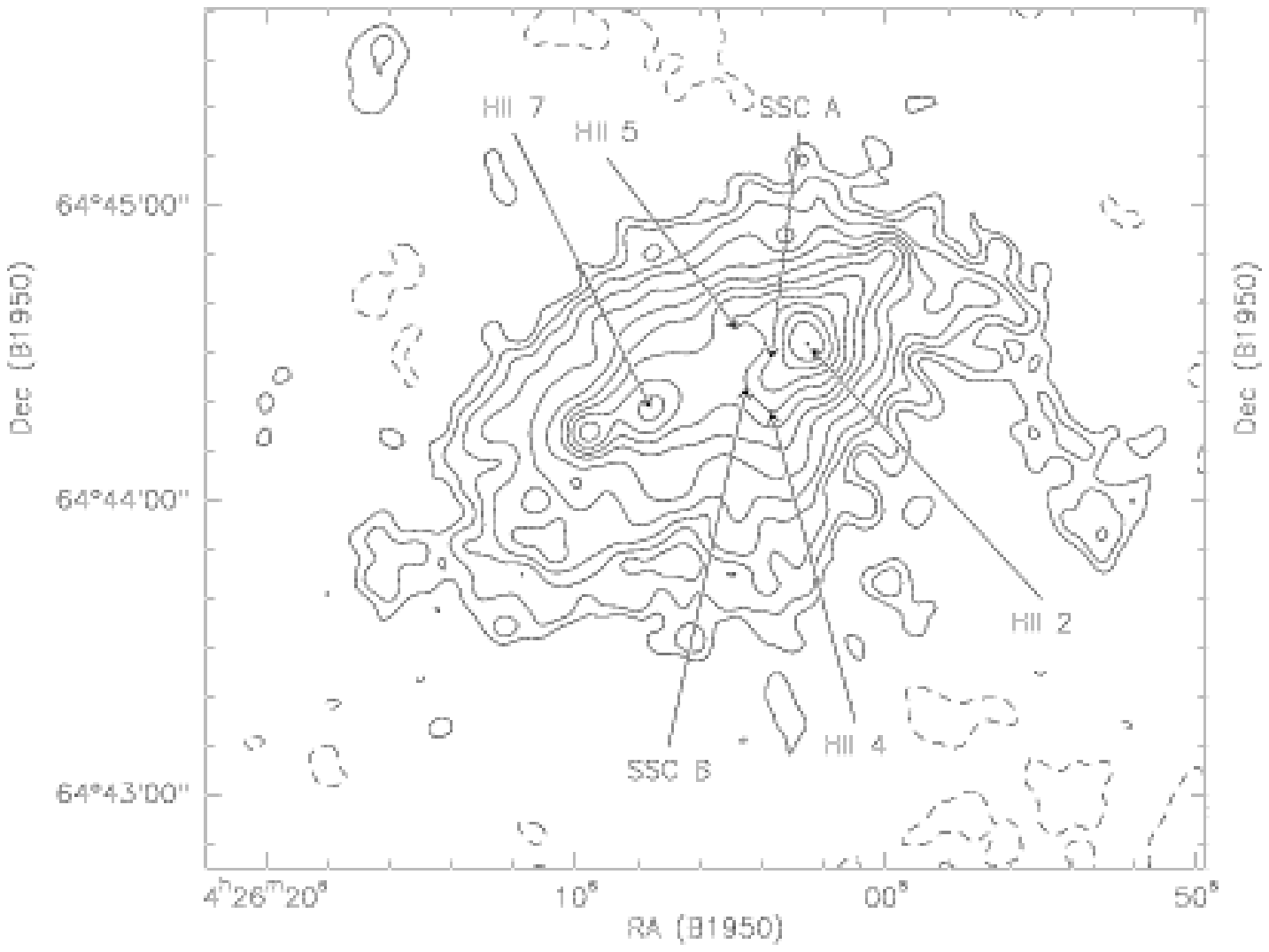}}}
{\rotatebox{270}{\includegraphics[width=6.5cm]{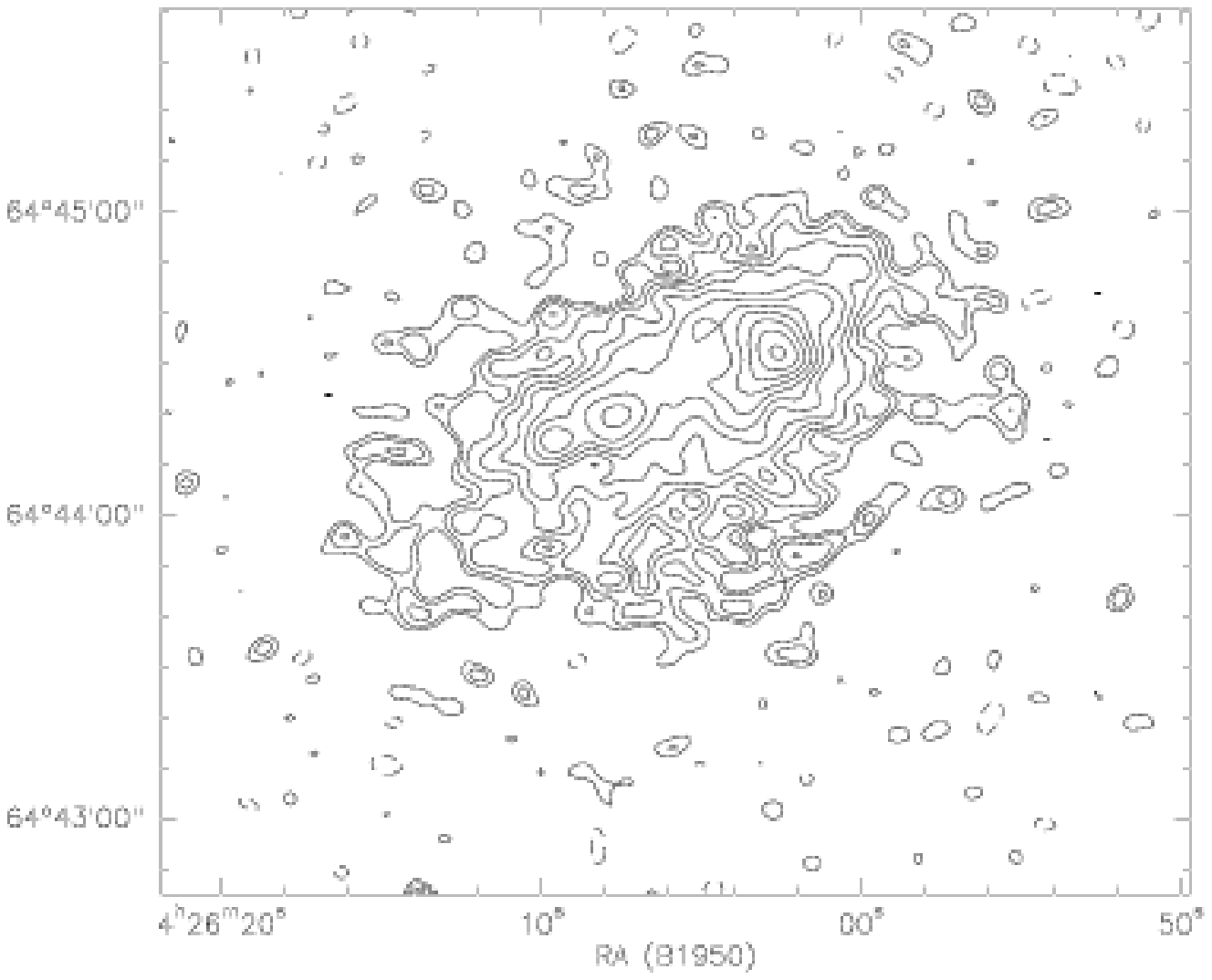}}}}
\vskip0.4cm
\centerline{\hoffset-0.5cm8.4~GHz\hskip8.5cm 15.4 ~GHz}
\hbox{
{\rotatebox{270}{\includegraphics[width=6.5cm]{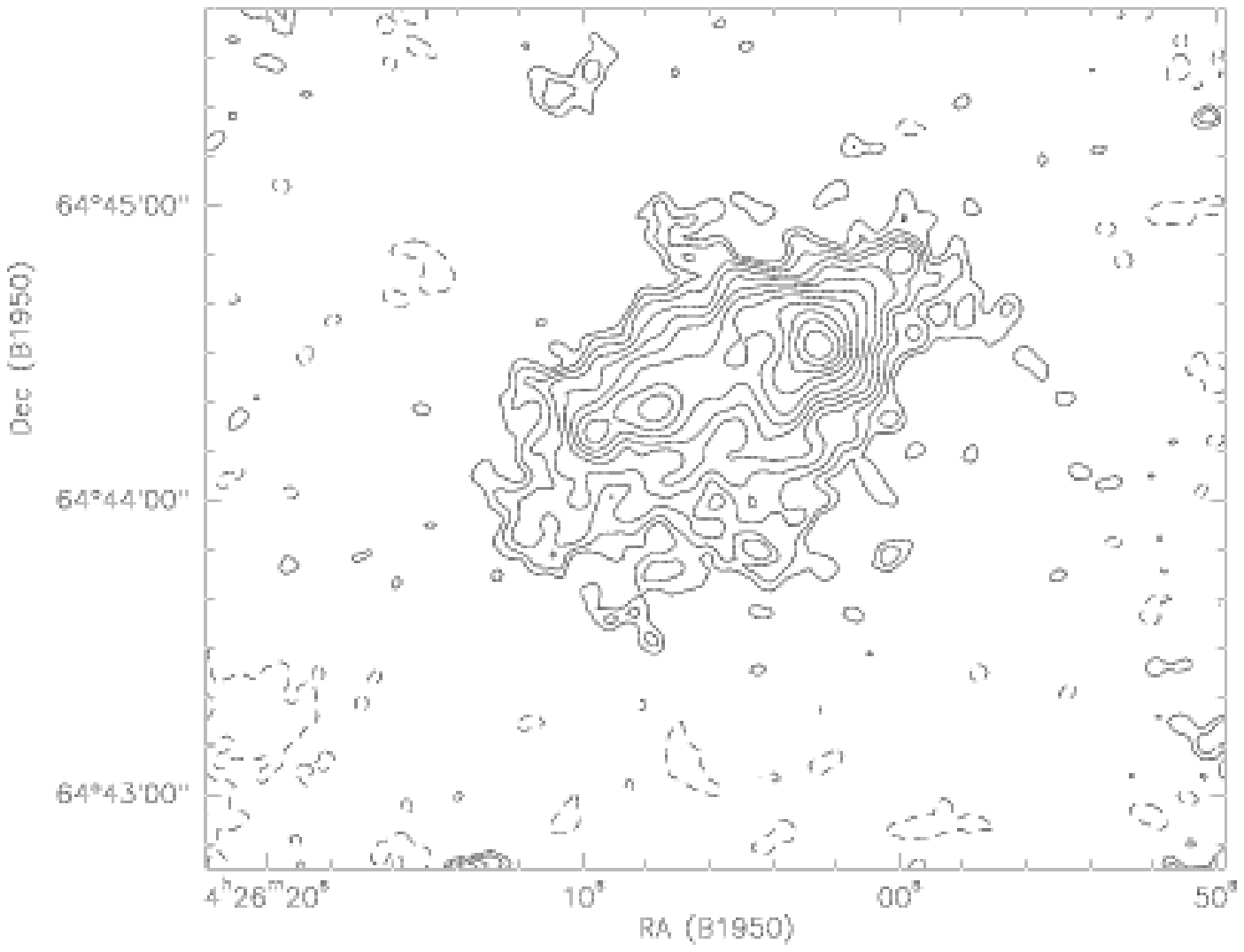}}}
{\rotatebox{270}{\includegraphics[width=6.5cm]{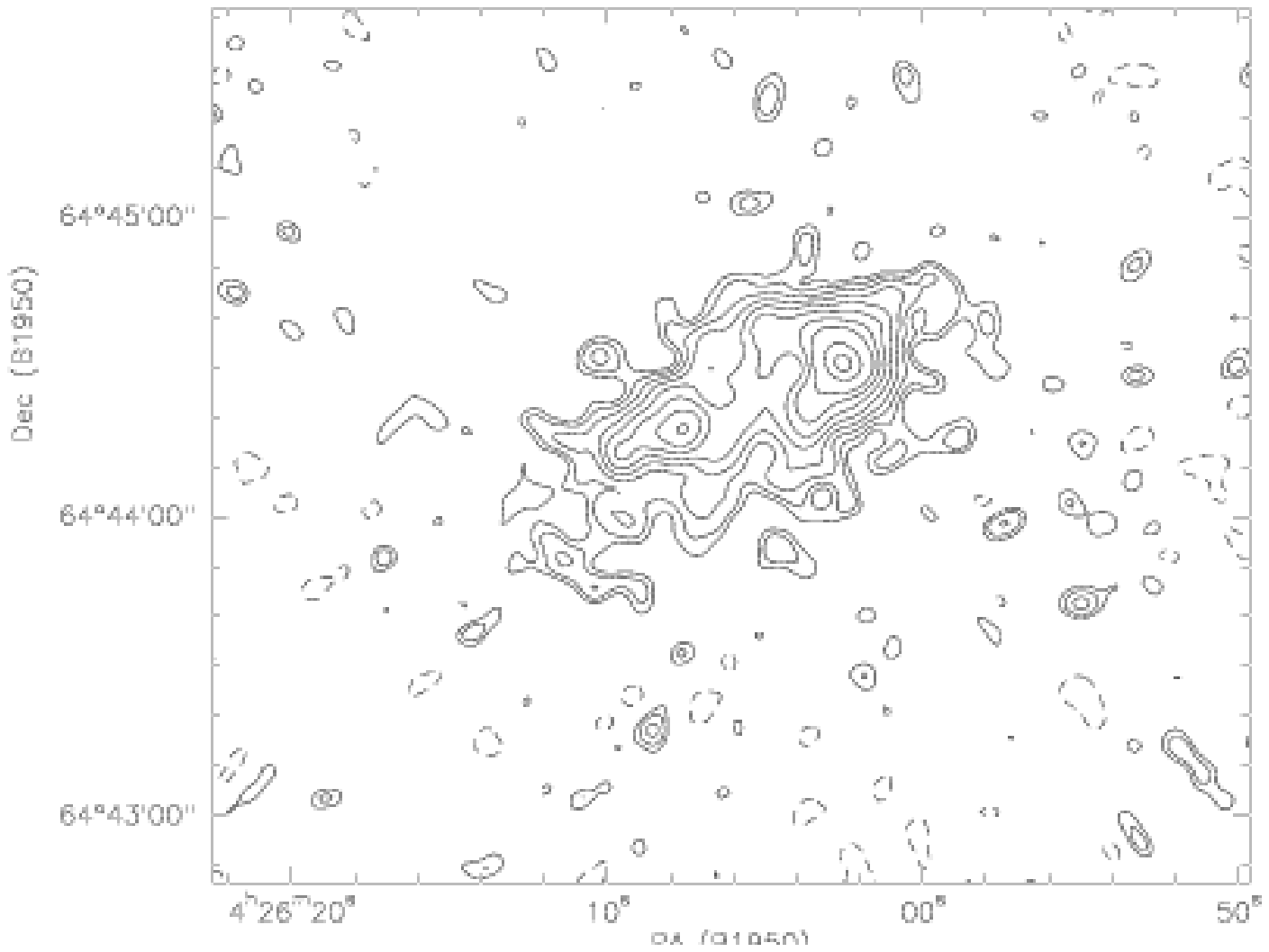}}}}
\caption[Maps of NGC~1569]{
Maps of NGC~1569 at 1.5, 4.9, 8.4 and 15.4~GHz.  The resolution is
$6.0 \times 6.0$~arcsec$^{2}$ for each map.
The contour levels start at two times the rms noise
(see Table 1), corresponding to 0.24~mJy/beam, 0.26~mJy/beam,
0.18~mJy/beam and 0.30~mJy/beam at 1.5, 4.9, 8.5 and 15.4 GHz, 
respectively, and increase by factors of $\sqrt{2}$.
In the 1.5~GHz map the positions of the strongest \hii \ regions
of \citet{wal91}  are shown, as well as the positions of the 
super star clusters (SSC) A and B.
}
%
\label{maps}
\end{minipage}
\end{figure*}

NGC~1569 was mapped with the VLA\footnote{
The VLA is operated by the National Radio Astronomy Observatory for
Associated Universities Inc., under a cooperative agreement with the
National Science Foundation.}
at 8.4~GHz and with the Cambridge
Ryle Telescope (RT) at 4.9 and 15.4~GHz.  Additionally, existing VLA data
(B-array) at
1.46~GHz \citep{con83} were available.
When the observations of NGC~1569 were made (between 1990 and 1993),
the RT had a band-width of
280~MHz split into 28 10-MHz frequency channels. Together with the
minimum baseline of 18~m this provides excellent temperature sensitivity
at both 4.9 and 15.4~GHz. The maximum baseline of the RT is 4.6~km.

The 1.5~GHz data were from \citep{con83} and no additional reduction
was required. This map was made with uniform weighting and a 
Gaussian taper in the {\it uv}-plane of $\sim$ 30000 $\lambda$,
resulting in a $\sim$ 6\arcsec synthesized beam. 

The 8.4-GHz VLA observations were in the B, C and D arrays and were
observed during 1990/91.  
3C286 and 3C48 were used as flux calibrators.
Reduction of the data followed standard VLA
procedures  in {\sc aips}. The data were self-calibrated to convergence 
using a model consisting of the positive {\sc CLEAN} 
components from the task CALIB. 

The RT observations were carried out between September 1990 and February 1993.
A total of five telescope configurations at each
observing frequency were employed giving a nearly fully filled aperture
out to a  baseline of about 1 km.
A minimum of two 12-hour runs in each
configuration were obtained (any observation badly
affected by interference was rejected).
This gave a resolution of approximately
$1.0 \times 1.0 \mbox{\rm cosec}( \delta )$~arcsec$^{2}$ at 15.4~GHz and
$3.0 \times 3.0 \mbox{\rm cosec}( \delta )$~arcsec$^{2}$ at 4.9~GHz.
At 4.9~GHz observations of phase-calibrators were made at the beginning
and end of each run, while at 15.4~GHz calibration observations were
interleaved with those of NGC~1569.  3C~286 and 3C~48 were observed
regularly as flux calibrators.
Calibration and data-editing (principally to remove narrow-band and
time-varying interference)
was performed in the MRAO package {\sc postmortem}, with subsequent reduction
in {\sc aips} and the MRAO package {\sc anmap}.

The maps at 4.9, 8.4 and 15.4 GHz were matched to the same spatial
sensitivity as the 1.5 GHz map by  restricting the {\it uv}-range
to values between $1000\lambda$ and $30000\lambda$, 
so that all maps were sensitive to spatial scales between about 6\arcsec
and 120\arcsec. 
After this, the maps were  {\sc CLEAN}ed.
All maps have been corrected for the primary beam responses of the
telescopes.
This set of four images is shown in Fig. 1 and a summary of their
characteristics is given in Table 1.

\begin{table}
\caption{Multifrequency radio maps}
\begin{tabular}{@{}lcc}
Frequency & Telescope & r.m.s  noise$^{(1)}$\\
 \  [GHZ]     &           & [mJy/beam] \\
\hline
1.5 & VLA (B-array)$^{(2)}$ & 0.12 \\ 
4.9 & Ryle                &  0.13 \\ 
8.4 & VLA (B,C,D-array)   &  0.09 \\
15.4 & Ryle               &  0.15 \\
\hline
\end{tabular}

(1) at a resolution of $6\times 6$ \arcsec (as the maps in Fig. 1)

(2) from \citet{con83}

\end{table}

The data at 1.5, 4.9 and 8.4 GHz are affected by the lack of short
baselines which is reflected as reduced integrated flux densities when
compared to single-dish measurements (see Tab. 2).
Only the map at 15.4 GHz,
where most of the emission is very concentrated, does
not show a considerable flux deficit.
In order to make the spectral index comparison as accurate as
possible, it is therefore
crucial to ensure that the spectral index is determined
from images all of which are sensitive as far as possible to the same
range of spatial structure. 
The B-array VLA map at 1.5 GHz had the
smallest range of spatial sensitivity and restricted therefore the
{\it uv}-range used in the other maps.
To achieve the match, the {\it uv}-range of the
data in the aperture plane used in the construction of the images above
1.5 GHz was restricted during the mapping process so that the
minimum and maximum baselines (measured in wavelengths) used in the
imaging were the same; this involved removing baselines from all the
high-frequency data. 
While it is not possible to ensure that the
precise sampling of the aperture plane is the same in all cases, for both
the VLA and RT the dense sampling near the shortest baselines
leads to similar overall coverage of the {\it uv} plane.

\begin{table}
\caption{Integrated flux densities}
  \begin{tabular}{@{}lccc}
Frequency & Flux density & Telescope & Reference \\
$ $ [MHz]     &  [mJy]        &           &           \\
\hline
38  & 2300 $\pm$ 580 &  CLFST &  1 \\
57.5 &1600 $\pm$ 500 & Clark Lake  &  2\\
151  & 940 $\pm$ 180 &  CLFST & 1 \\
610        & 610 $\pm$ 20 & WSRT &  3 \\
1415        & 440 $\pm$ 15 & WSRT & 3  \\
1415       & 410 $\pm$ 25 &  WSRT  & 5 \\
1465       & 285 $\pm$ 30 & VLA  & 6,16 \\
1490       & 411 & VLA & 7\\
2695      &  270 $\pm$ 50 & Effelsberg  & 8\\
2700      & 365 $\pm$ 23 & Green Bank & 9\\
4750      & 263 $\pm$ 20 & Effelsberg & 10\\
4850      & 202 $\pm$ 19 & Green Bank & 11,12\\
4919       &  210 $\pm$ 20 & Ryle  & 6 \\
4995        &  280 $\pm$ 20       & WSRT & 3 \\
5000      & 276 $\pm$ 42 & Green Bank & 9\\
6630      & 235 $\pm$ 30 &  Algonquin  & 13\\
8415       &  125 $\pm$ 12 & VLA  & 6 \\
10700      & 156 $\pm$ 16 & Effelsberg & 10 \\
10700      &  154 $\pm$ 4 & Effelsberg & 14\\
10700      &  171 $\pm$ 10 & OVRO  & 15\\
15360       &  116 $\pm$ 13& Ryle  & 6 \\
24500      & 96 $\pm$ 8 & Effelsberg & 10\\

\end{tabular}
\\
The tables includes the data listed in Israel \& de Bruyn (1988)
(except for the points that they rejected as unreliable)
as well as more recent data.
References:
(1) \citet{how90}, 
(2) \citet{isr90}, 
(3) \citet{isr-deb88}, 
(4) \citet{con98},  
(5) \citet{hum80}, 
(6) this work,
(7) \citet{con87},  
(8) \citet*{pfl80}, 
(9) \citet{sul76}, 
(10) \citet{kle86},  
(11) \citet{gre91}, 
(12) \citet*{bec91},
(13) \citet{mcc73},
(14) \citet{nik95}, 
(15) \citet{isr83}
(16) \citet{con83}
\end{table}
\begin{figure}
{\rotatebox{270}{\includegraphics[width=8.cm]{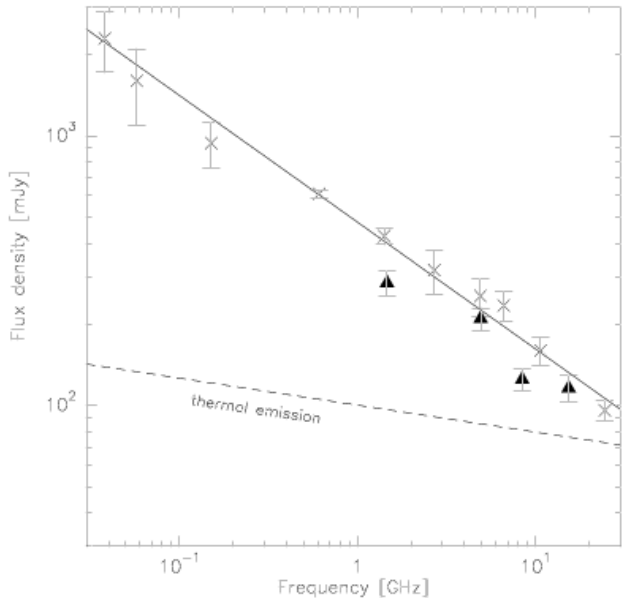}}}
\caption{Radio spectrum of NGC~1569. The data points are taken
from Table 2.  The triangles show our data and
the crosses the data of other authors
(shown as averages for observations at similar frequencies).
The full line shows a single-power law fit with $\alpha=0.47$ to the data
and the dashed line the thermal radio emission derived from the
spectral fitting (see Section~4).
}
\label{spectrum-data}
\end{figure}

\section{The Radio Properties of NGC~1569}

\subsection{Structure of the radio emission}

Figure~\ref{maps} shows the radio emission at the four frequencies at a
common resolution of 6 arcsec.
At 1.5~GHz most of the
emission lies beyond the optical galaxy in an irregularly-shaped halo.
Prominent is an arm of emission to the West of the galaxy which
coincides with a similar structure in H$\alpha$ \citep{wal91}.
Along the optical disk the radio emission shows three
(at 15~GHz only 2) peaks. The two peaks visible at all four
wavelengths coincide with the two brightest \hii \ regions found
by \citet{wal91} (number 2 and 7 in his nomenclature).
Apart from these features, the overall distribution of the
radio emission is in general very
similar to the \halpha \ map of \citet{wal91}.
At higher frequencies the emission becomes more concentrated
towards the major axis.

\subsection{Spectral index map}
\begin{figure}
{\rotatebox{270}{\includegraphics[width=6.cm]{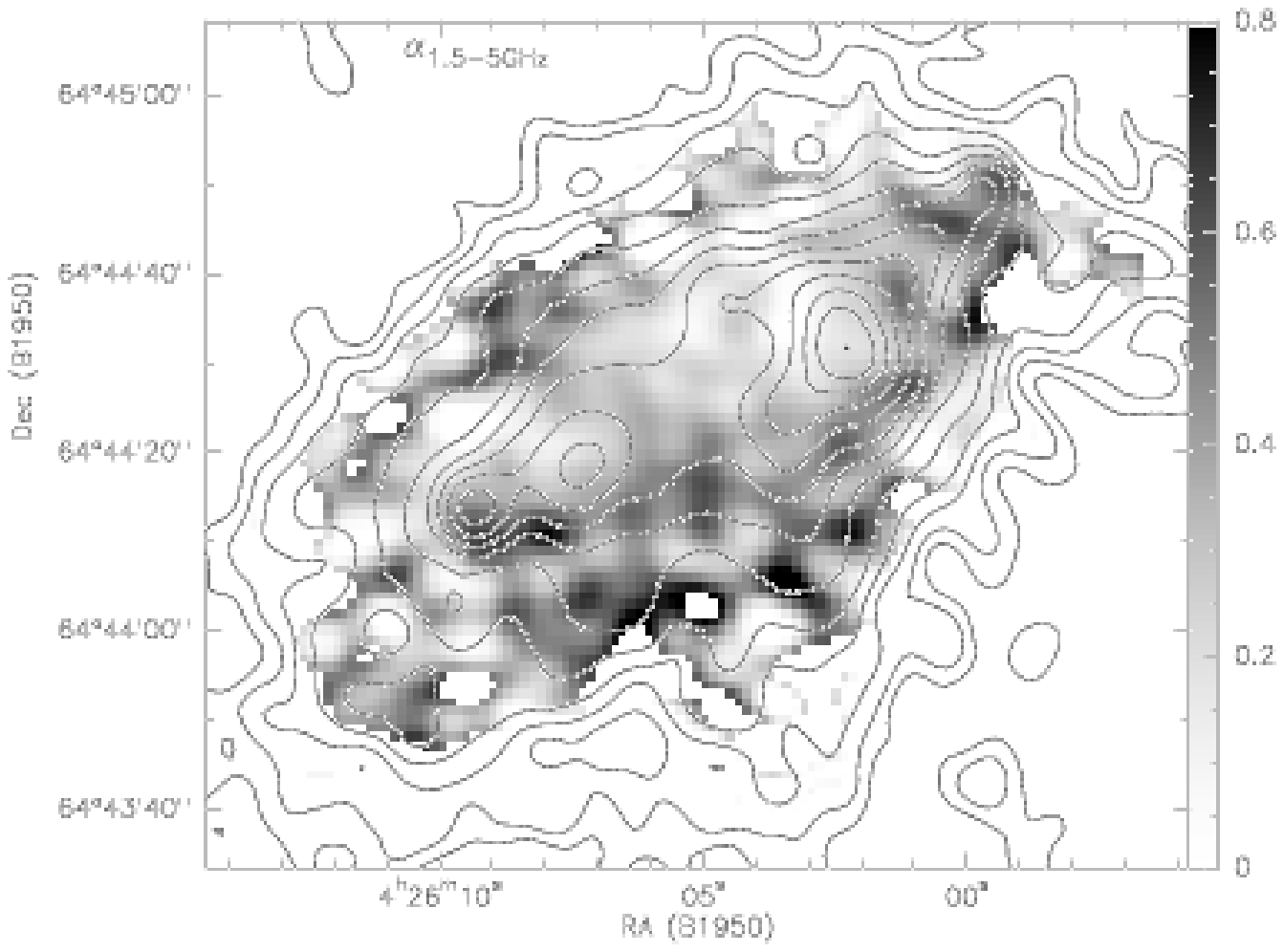}}}
{\rotatebox{270}{\includegraphics[width=6.cm]{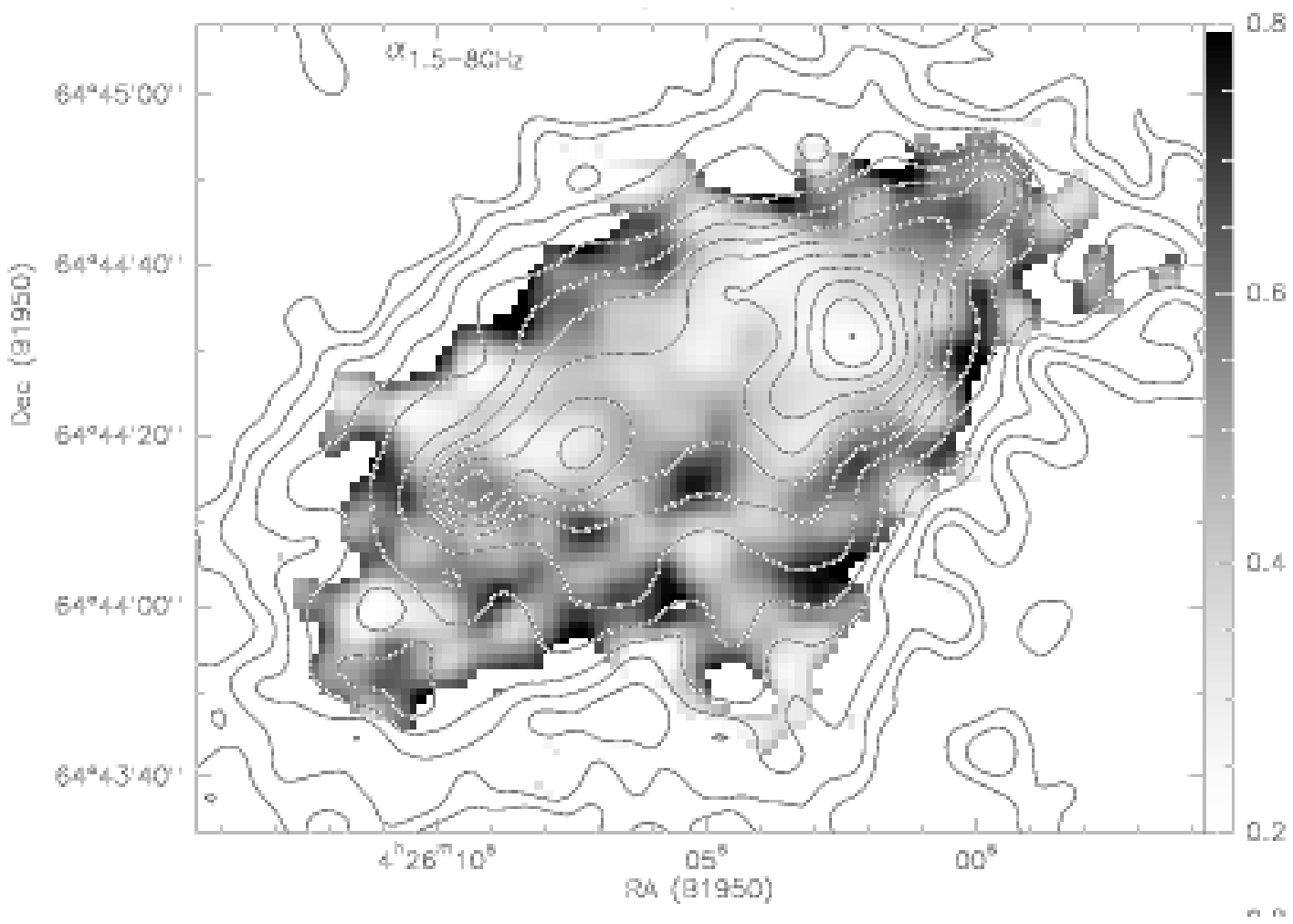}}}
{\rotatebox{270}{\includegraphics[width=6.cm]{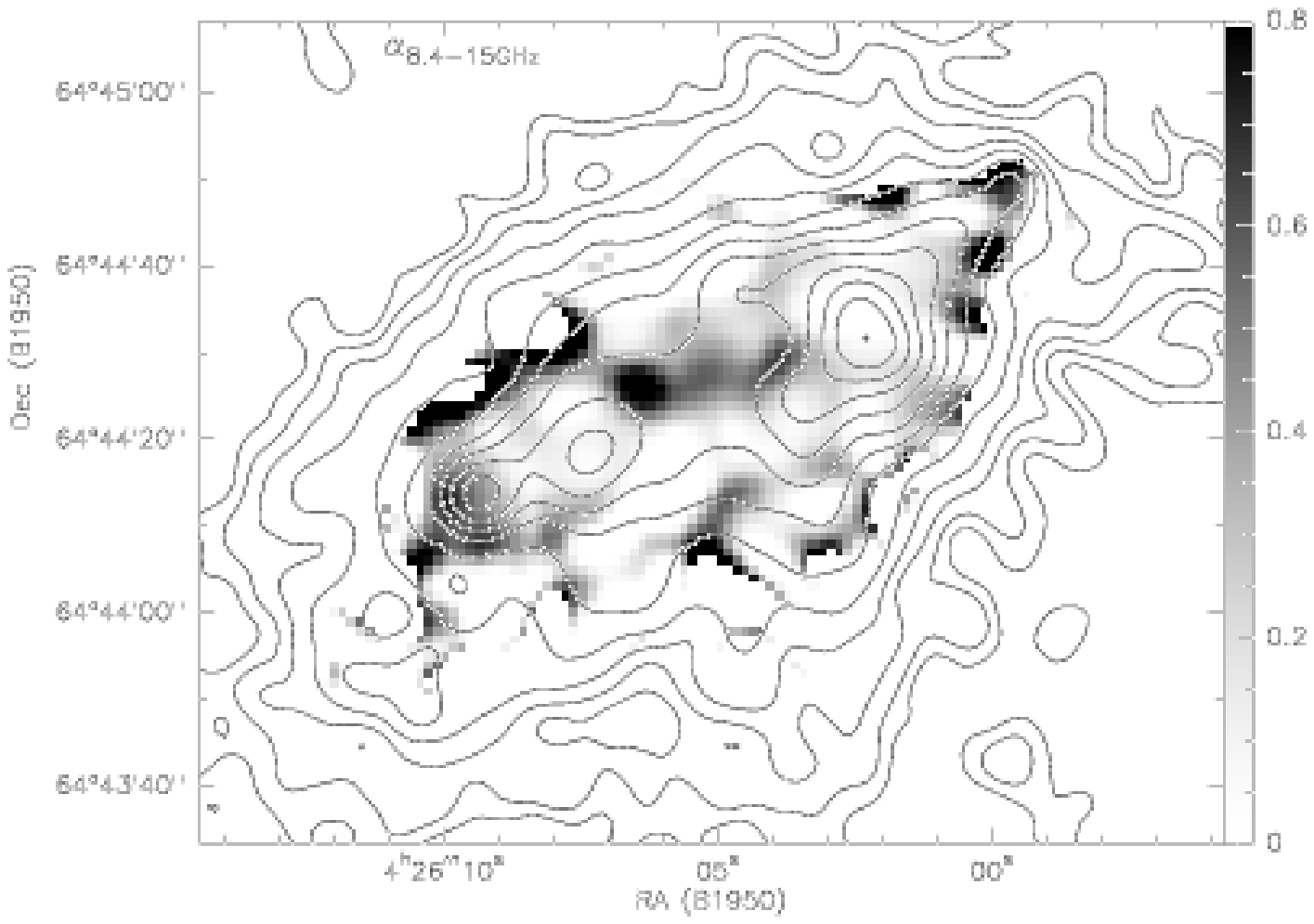}}}
\caption{
Spectral index distribution between 1.5 and 5 GHz (top),
1.5 and 8.4~GHz (middle) and 
8.4 and 15.4~GHz (bottom) at a resolution of
6.0$\times$6.0~arcsec$^{2}$, overlaid over contours
of the 1.5 GHz emission (contour levels as in Fig. 1). 
The spectral index is calculated for
all pixels where the flux exceeds three times the r.m.s. noise.
The greyscales range from 0.0 to 0.8 ($\alpha_{\rm 1.5-5 GHz}$
and $\alpha_{\rm 8.4-15GHz}$) and from
0.2 to 0.8 ($\alpha_{\rm 1.5-8.4GHz}$).
}
\label{spectral-index}
\end{figure}
The spectral index maps between 1.5 and 4.9 GHz, between 1.5 and 8.4 GHz
and between 8.4 and 15.4~GHz
are shown in Fig.~\ref{spectral-index}. Most striking  is the extreme
patchiness of these maps.
The spectrum around the two main maxima of the radio emission
and along the disk is flat and there is a trend for the
low-frequency spectrum to steepen away from the galactic disk,
up to values of $\alpha = 0.9$ at
the outermost end of the halo. This steepening
does however not occur everywhere and not in a regular
way. There are numerous filamentary structures where the
spectral index is much lower than in its surroundings and
 at some places  the spectrum even flattens with
increasing distance from the centre.
The high-frequency spectrum ($\alpha_{\rm 8.4-15GHz}$
shows an elongated region in the centre of the galaxy
with a steep spectrum indicating the predominance of synchrotron emission
in this region.

\section{Spectral Model fitting}

The radio-continuum emission consists of both a non-thermal
(synchrotron) component and thermal emission.  We use a spectral-fitting
technique (the same as in  \citet{lis98}, and similar to that
in \citet{cle99}),
to estimate the two contributions and to determine spatial
variations in the form of the synchrotron spectrum.
For the set of maps of Fig. 1 we fit a function of the following
form at all pixels where the flux densities at the four frequencies
exceed three times the noise:
\begin{equation}
I(\nu)=S_0\, C(x_{\rm B},\gamma)+T_0\nu^{-0.1},
\end{equation}
where $T_0\nu^{-0.1}$ is the form of the spectrum from an
optically thin \hii region, $S_0 C(x_{\rm B},\gamma)$ is the
synchrotron spectrum from an aged electron population with injection index
$\gamma$ which have been radiating for a time $t$ and
$x_{\rm B}=\nu/\nu_{\rm B}$, where $\nu_{\rm B}$ is the break frequency
(see below).
During the fitting procedure the minimum in $\chi^2$ is found by
varying $S_0$, $\nu_{\rm B}$
and $T_0$ with the additional constraint that they are
positive definite.

The form of the synchrotron spectrum $C(x_{\rm B},\gamma)$
depends on many processes, but to
simplify the problem we make the following approximations:
\begin{enumerate}
\item
The dominant electron energy loss mechanism is by synchrotron radiation
and inverse Compton losses.
These
energy losses are proportional to the energy density of the magnetic
field, $\ub$, and the energy density of the radiation field,
$\urad$, respectively.
The magnetic field can be estimated from the standard minimum energy
assumption. With a radio size of $3.5'$, a high-frequency cut-off
of 8~GHz, a low-frequency cut-off of 100~MHz and a spectral index of
0.45, this yields $B=12\mu$G (Israel \& de Bruyn 1988).
$\urad$ can be estimated from the bolometric luminosity, $\lbol$,
of the galaxy which
we approximate here as the sum of the infrared, optical and UV emission taken
from \citet{wal91}, ($\lbol=7.4 \times10^{35}$ W), by assuming that the
radiation emitting matter is distributed spherically:

\begin{equation}
\urad={L_{\rm bol} \over c\pi R^2}=
{\bigl(L_{\rm bol}/ {\rm W}\bigr) \over \bigl(R/{\rm kpc}\bigr)^2}
\times 1.4 \times
   10^{-35}\biggl[{\rm eV \over cm^3}\biggr].
\end{equation}
We derive, with $R=1.7$ kpc (the average of major and minor
axes of 3.6 and 1.8 arcmin), $\urad=3.6$ eV cm$^{-3}$.

\item We neglect propagation effects within the galaxy.
A spatially resolved model, as has been done
for  NGC~2146 \citep{lis96},
is not possible because of the extreme complexity of this galaxy
making it impossible to derive the
distribution of the sites of CR
acceleration as a function of time.
Given our small beam size (6 arcsec corresponds to only 66 pc)
this assumption most likely does
not represent reality, because CREs are able to
diffuse this distance (assuming a diffusion coefficient of
$10^{29}$ cm$^2$ s$^{-1}$ as in our Galaxy) in about $10^4$ yr,
a time much shorter than the energy loss time of typically
$10^7$ yr.
However, this simplifying assumption still allows us to infer a
realistic picture of the age distribution of the CREs,
because the propagation in NGC~1569 is to a large extent
driven by convection, as argued
in Sect. 5 and 6, and thus by an  energy-independent
process, which does not alter  the shape of the synchrotron spectrum.
Therefore, we expect that the CRE spectrum in each pixel
mainly  reflects the energy losses.

\item  We assume that
the electrons which are radiating within a given resolution element
were all accelerated (injected) a time $t$ ago (the electron age).
An alternative hypothesis for the star formation history was tested by
\citet{cle99}. They adopted in addition to the present
star formation history a
continuous injection of CRs starting some time $t$ ago. The results
for both star formation histories were  practically  identical.

\item The electron loss time is much greater than the pitch angle randomization
time --- the so-called JP model \citep{jaf74}.

\item The electron injection index is 1.8, close to the value
of 2.0 predicted from simple theoretical models of shock acceleration,
and matching the observed low-frequency spectral index of NGC~1569.

\end{enumerate}

With these assumptions the electron energy distribution with an
injection index $\gamma$ is given by:
\begin{equation}
N(E,\theta,t) = \left\{ \begin{array}{ll}
N_{0}E^{-\gamma} (1 - \epsilon (\ub+\urad) E t)^{\gamma -2} & E < E_{\rm B} \\
0 & E \ge E_{\rm B} \\
\end{array} \right.
\end{equation}
with
$\epsilon=(4/3)\sigma_{\rm T}/(m_{\rm e}^2 c)$,
where $\sigma_{\rm T}$ is the Thomson cross-section,
and $E_{\rm B}=[\epsilon(\ub+\urad)t]^{-1}$ is the break energy.
The synchrotron spectrum is then obtained by convolving this
electron energy distribution with the spectrum of the
synchrotron emission of a single electron
(see \citet{lis96} for explicit formulae).
The break frequency is given by:
\begin{equation}
\nu_{\rm B} =(3/2)\nu_{\rm G}E_{\rm B}\propto \frac{B}{(\urad+\ub)^2t^2}
\end{equation}
where $\nu_{\rm G}$ is the gyrofrequency.

\subsection{Results}
%

\begin{figure}
{\rotatebox{270}{\includegraphics[width=6.cm]{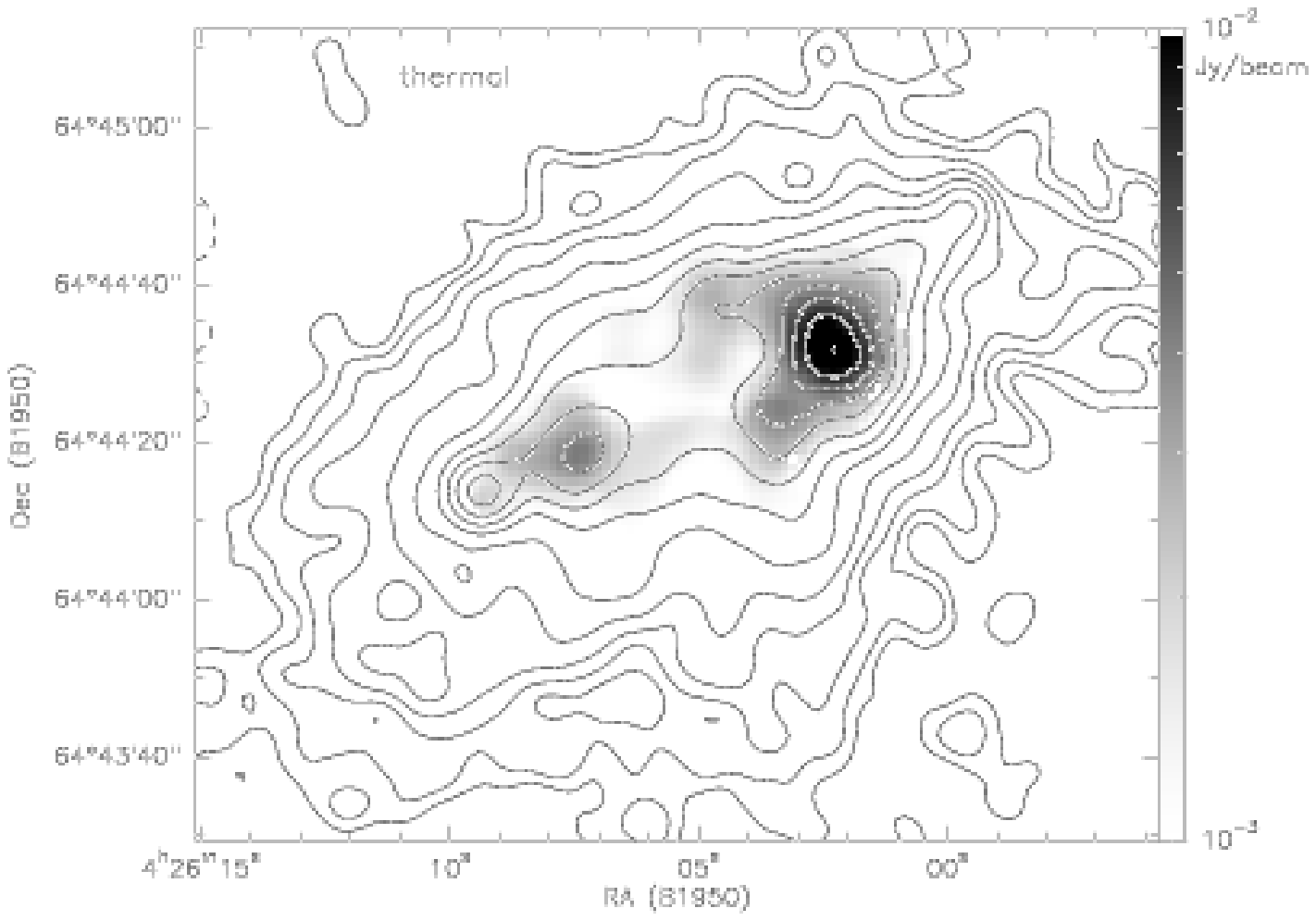}}}
{\rotatebox{270}{\includegraphics[width=6.cm]{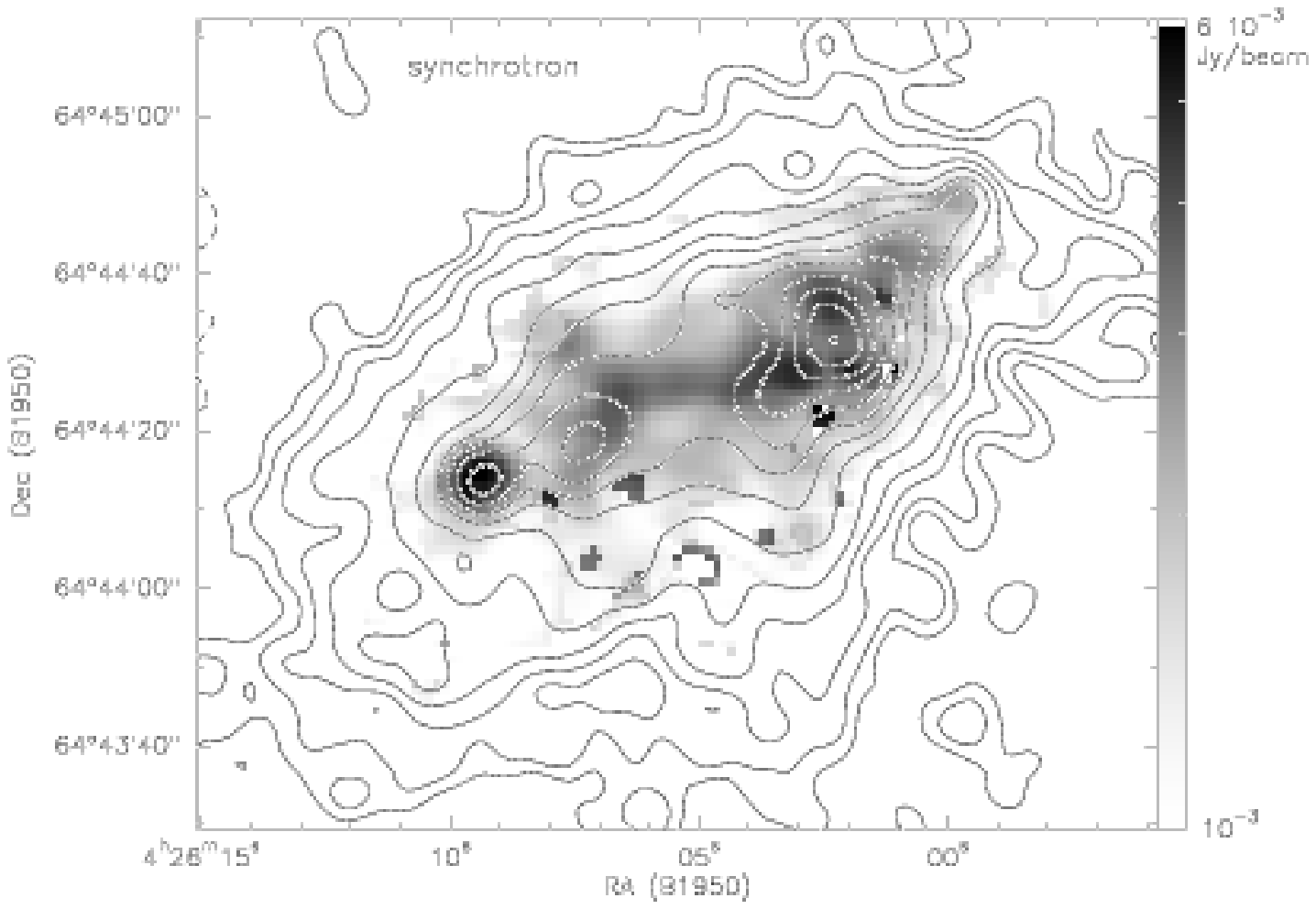}}}
{\rotatebox{270}{\includegraphics[width=6.cm]{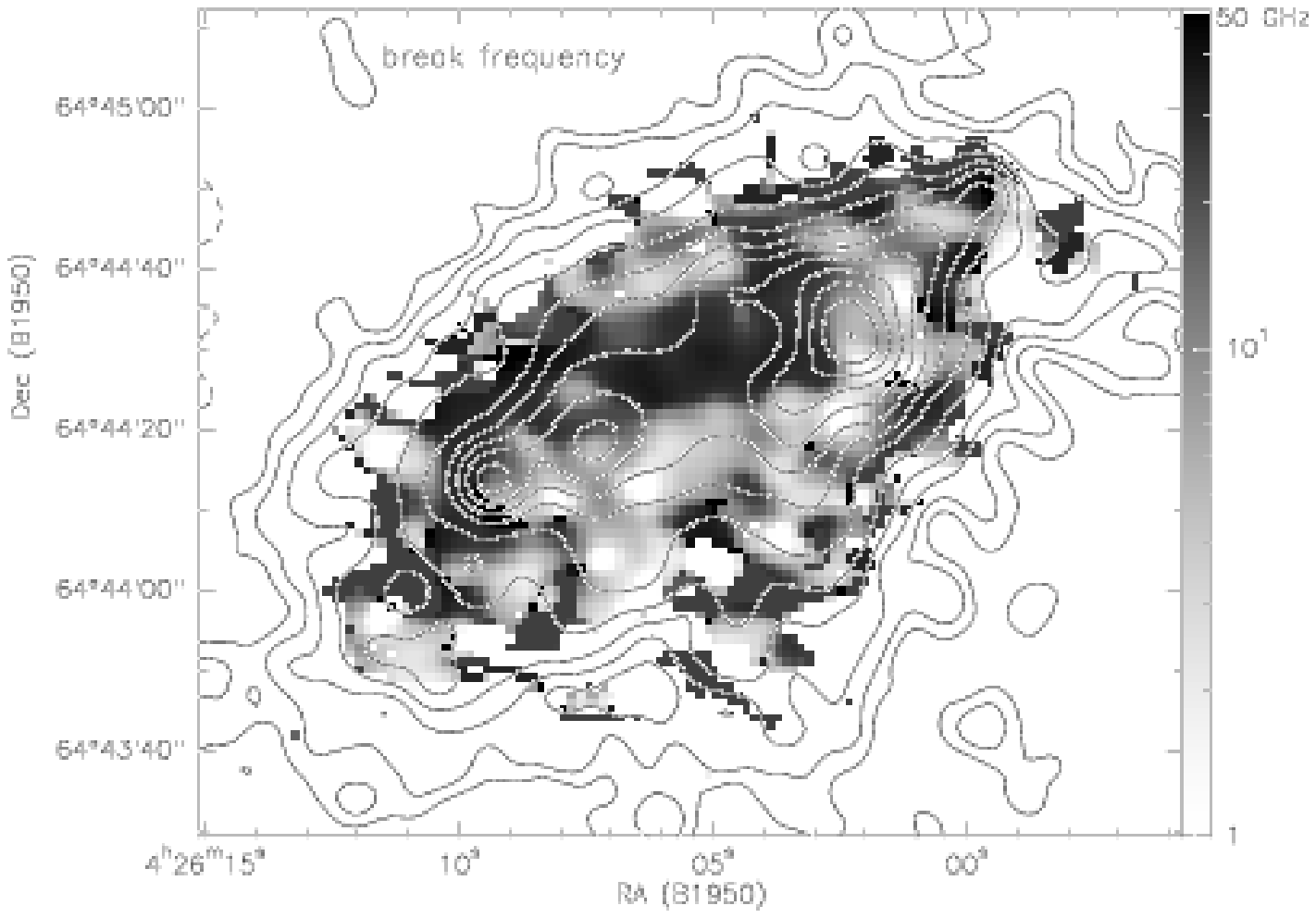}}}
\caption{Greyscale maps of NGC~1569, showing
the thermal emission (top),
synchrotron emission (middle), both at 1 GHz,
and break frequency (bottom)
derived from the break in the radio spectrum,
overlaid over contours of the 1.5 GHz emission.
The greyscales are logarithmic in order to better highlight
structures at different intensities.
The range of break frequencies from 1 (white) to 50 GHz (black)
corresponds to electron ages from
 $1.4 \times 10^6$ (black) to $1.0\times 10^7$ yr (white)
for the adopted parameters of $B$ and $\urad$.
}
\label{therm-syn-break}
\end{figure}
%
Maps of the synchrotron and thermal emissions and break frequency (which
can be translated to CR electron age)
are shown in Fig. \ref{therm-syn-break}.

The integrated flux density of the
thermal emission (Fig. \ref{therm-syn-break} top) that we derive at
1 GHz is $\sim 100$~mJy. This number should be unaffected by
the missing extended flux of our interferometer observations because
the thermal emission is very spatially concentrated as can be seen
from the 15.4~GHz map in Fig.~\ref{maps}.
The total thermal radio emission that we derive  by our
spectral fitting method corresponds very closely to the
value of 97~mJy at 1~GHz
estimated by \citet{isr-deb88}
from reddening-corrected \halpha \ measurements.
Most of the thermal radio emission is concentrated in the maximum
of the radio continuum emission and coincides with the strongest
\hii \ region (number 2) of \citet{wal91}. Correspondingly,
the second strongest \hii \ region
of Waller (number 7) is the second peak of the
thermal radio emission. This good agreement between the \halpha \
and thermal radio emission provides further support to the  validity of our
fitting method.

The synchrotron emission (Fig. \ref{therm-syn-break} middle)
shows only one pronounced peak at the south-eastern end,
corresponding
to the second strongest peak of the 1.5 GHz emission.
Interestingly, this peak is not centered in the galaxy
 and is far away from the
major sites of star formation traced by the thermal radio emission
or the SSCs A and B.
In the rest of the disk, the synchrotron emission is concentrated towards
a bar-like structure along the major axis extending
 in between the positions of the SSC A and B.
The peaks of the thermal and nonthermal
radio emission do not coincide. At the maximum of the thermal emission
there is only a moderate amount of synchrotron emission but
surrounding this position there are several local maxima of
synchrotron emission.

The map of break frequency (Fig. \ref{therm-syn-break} bottom) gives ages
of several $10^6$ years. The distribution is very patchy and
can be divided approximately into two zones, one with relatively
young CREs (ages of $\sim 1\times 10^6$ years) and other regions
with older CREs  (ages of $\sim 1\times 10^7$ years).
The young CREs coincide with the main synchrotron emission
along the major axis of the galaxy, and extend
slightly north of it. Towards the halo, the electrons
generally become older
although there remain patches of relatively young CREs.

The large frequency range used  -- 
from 1.5 GHz where synchrotron emission contributes
about 75\% to the total emission up to 15.4 GHz where the thermal
emission is the dominant ($\sim$ 90\%) contribution -- allows
a reliable separation of both emissions based on their different spectral 
shapes. We therefore expect the error in the distribution 
of the synchrotron and thermal emission
to be small and in the range of the errors of the
individual maps.
This is illustrated by a comparison of Fig. 3 and 4 
which shows that the thermal emission is strong
a places with a flat spectrum whereas strong synchrotron emission 
coincides with a steep spectrum  up to high frequencies. 
The determination of the break frequency can be much less reliable
at places with dominant thermal and
little synchrotron emission
 and at those places where the synchrotron
spectrum shows no clear curvature, leading to values of the break frequency
below 1.5 or above 15~GHz. Such values outside the range covered by
our radio data can only be estimates.
The calculation of the electron age from the break frequency depends 
directly on the adopted values for the magnetic and radiation field.
Spatial variations of these two quantities, which are expected 
to be present but are not
included in our model, produce therefore
an error in the electron age.
The separation of the thermal and synchrotron emission and the
determination of the break frequency are however not affected by
uncertainties in the magnetic and radiation field.

\section{Total synchrotron spectrum}

\begin{figure}
{\rotatebox{270}{\includegraphics[width=6.5cm]{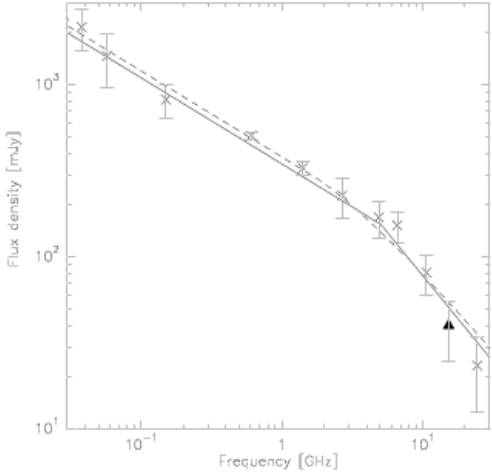}}}
\caption{The integrated synchrotron spectrum, obtained from the
reliable data points in Fig. \ref{spectrum-data}
by subtracting of a thermal radio emission of
$P_{\rm therm}=100 \times (\nu/{\rm GHz})^{-0.1}$ mJy together with
two fits to the data: time-dependent star-formation (dashed line)
at $t=1\times 10^7$ yr
(see Section 5.1 for details of the model)
and convection into an open halo (full line,
the model is described in Section 5.2).
}
\label{spectrum-fit}
\end{figure}

The thermal radio emission derived from the spectral
fitting procedure of the radio maps allows us
to determine the total thermal radio emission and, by subtraction,
the total synchrotron spectrum (Fig.~\ref{spectrum-fit}). 
Our result confirms the
conclusion of \citet{isr-deb88} that the
synchrotron spectrum shows an abrupt break, clearly visible in the
data points above 10~GHz, because our derivation of the thermal
radio emission coincides very well with the estimate of 
\citet{isr-deb88} from extinction-corrected H$\alpha$ emission.

The shape of a synchrotron spectrum is determined by the energy losses,
CR propagation processes and the star formation history.
We have modeled various possibilities that could produce such a
break and found that only two of them
are in principle
able to explain it. In the next subsections we will describe
them as well as the other processes considered.

\subsection{Time-dependent star formation}

A temporal change in the SFR produces a change in the SN rate and
therefore in the CRE production.
If the time-scale of this change is  shorter than the
energy loss time scale of CREs it will be noticeable
in the synchrotron spectrum. \citet{dee93} have found
evidence for this mechanism to be important in shaping the
spectrum of several \hii galaxies.
NGC~1569 is a dwarf galaxy with variable SF in the recent past
which can be expected to be
reflected in its synchrotron spectrum.

The break in the synchrotron spectrum is very sharp, therefore
we will assume a change in the SFR as abrupt as possible.
We assume that SF
started at $t=0$ and ended at $t_{\rm end}$.
The exact value of  $t_{\rm end}$ is not
important, as long as it is sufficiently large
(here we chose $t_{\rm end}=10^8$ yr) to have achieved a
steady-state situation for massive stars and CREs by that time.
We further
assume that the stars are produced according to a Salpeter-IMF and that
all stars above 8 $\ms$ end their lives in a SN explosion which
accelerates CRs. As in the spectral fitting,
the main energy losses are assumed to be
synchrotron and inverse Compton
and we use the values for the magnetic field $B$ and the radiation energy
density derived in the previous section.

The synchrotron emission is then calculated by solving the
time-dependent equation for the electron particle density
$f(t,E)$:

\begin{equation}
{\partial f(t,E)\over \partial t}
        = \biggl({E\over mc^2}\biggr)^{-\gamma} q_{\rm SN}\nu_{\rm SN}(t)+
    {\partial\over \partial E} \biggr\{b
     \, E^2 \, f(t,E)\biggr\}.
\end{equation}

This equation takes into account  CR acceleration
in SN remnants with a source spectrum as
$(E/mc^2)^{-\gamma} q_{\rm SN}\nu_{\rm SN}(t)$
(where  $\nu_{\rm SN}(t)$ is the supernova
rate, calculated from the above defined SFR(t), 
and $q_{\rm SN}$ is the number of relativistic electrons 
produced per supernova and per unit energy interval), and the
subsequent radiative energy losses of the CR electrons
($b\propto (\urad+\ub)$).

The solution of Eq. (5) is given by:

\begin{equation}
f(E,t)=
\biggl({E\over mc^2}\biggr)^{-\gamma}
  q_{\rm SN}
\int^t_{t_0} \diff t'\,
\nu_{\rm SN}
(t')
 \biggl( 1-E\,b (t-t')\biggr)^{\gamma-2},
\end{equation}
where
$t_0 = max(0, t-\tauloss)$ and
$\tauloss(E,t)=(E\,b)^{-1}\propto ((\urad+\ub)E)^{-1}$ is the life-time
of the CREs against radiative energy losses.
Finally the total synchrotron spectrum is obtained by convolving $f(E,t)$
with the synchrotron emission  spectrum of a single
electron.

We calculated the spectra as a function of time and selected the best-fit
spectra. There are two possible times when a break in the
spectrum is produced:

1) A good fit to the data can be achieved soon after the
beginning of the starburst, at $t=1 \times 10^7$ yr
(see Fig.\ref{spectrum-fit}),
with a realistic
injection spectral index $\gamma=2.0$. The break in the spectrum
divides the low frequency range,
in which the electrons have not suffered
energy losses yet, from the high frequencies
where the spectrum
is steepened by 0.5 as a result of energy  losses. The position of the break is
given by the life-time of the electrons against radiative losses,
which is  $ 5\times 10^6$ yr at 5  GHz (roughly the
position of the break) with our values
for the magnetic field and $\urad$.

2) \citet{isr-deb88} suggested that the break in the
synchrotron spectrum was due to the abrupt {\it end} of the CR injection
about $5\times 10^6$ yr ago. A cut-off at high energies
is then produced above which
electrons have already lost their energy
since the time of the last CR injection.
We have tried to fit the data for times after
the end of the starburst.
Although a break is indeed produced by this
mechanism,
we could not find a satisfactory fit to the data.
Two reasons are responsible for this failure:
1) A visible break cannot be produced at the high frequencies where it
is observed.
The main reason is that,
 even for an abruptly
ending SFR, the SN rate does not cease immediately. If stars above
8 $\ms$, with life-times of $3\times 10^7$ yr, become SN, then
it takes about this time until the SN rate has ceased after the
end of the SF. This time-scale is
longer than the CR loss-time scale of $\sim 5\times 10^6$ years of CR electrons
emitting in the range around 5~GHz where the break is observed, so that
these high-energy CREs do not experience an
abrupt end in the injection.
If the SN rate ended abruptly, the break in the spectrum would be
more pronounced in better agreement with the data.
2) The low-frequency spectrum is very
flat, requiring an unrealistically low injection spectral index
($\gamma=1.0$) to explain it.

\subsection{Escape from the halo}

An alternative mechanism causing a break in the spectrum is escape
of CREs from the galaxy.
If the escape time-scale is comparable to the  energy loss
time-scale $\tauloss$ a break
of $\triangle \alpha = 0.5$ is produced in the spectrum which is
sufficient to explain the data if the break is sharp enough.
Convection as the mode of propagation is required for this.
If diffusion is the dominant propagation mechanism, the
frequency variation of the
expected radio spectral index is
too smooth to account for the sharp break in the
data (see next section).

We model this situation as follows: CREs that are produced in the
main disk of NGC~1569 convect into the halo with a constant
convection velocity, $V_c$, experiencing synchrotron
and inverse Compton losses at the same time. When they reach a distance
$z_{\rm halo}$ from the disk, they  can escape freely.
The one-dimensional convection-energy loss
equation describing this situation is:
\begin{equation}
{\partial f(E,z)\over \partial z} V_c -
{\partial\over \partial E} \biggr\{b\, E^2 \, f(E,z)\biggr\}=
\delta(z) \biggl({E\over mc^2}\biggr)^{-\gamma} q_{\rm SN}\nu_{\rm SN}.
\end{equation}
with $\delta(z)$ being the one-dimensional $\delta$-function.
The solution of this equation is given by:
\begin{equation}
f(E,z)= \frac{q_{\rm SN}\nu_{\rm SN}}{2V_c}
\biggl({E\over mc^2}\biggr)^{-\gamma}
\biggl(1-{bEz\over V_c}\biggr)^{\gamma-2}
\end{equation}
The synchrotron emission is calculated by first integrating this
emission from $z=0$ to $z0min(z_{\rm max},z_{\rm halo})$,
where $z_{\rm max}=V_c/(bE)$ is the maximum distance that CREs can
travel before having lost their energy to below $E$, 
and then convolving the result with
the synchrotron emission spectrum of a single electron.

We use the above values of $\urad$ and $\ub$, and estimate
$z_{\rm halo}=0.7$ kpc from the extent of the radio emission at
1.5 GHz. Then, we obtain a good fit to the data with
 $V_c=150$ km s$^{-1}$ (Fig.~\ref{spectrum-fit}).

\subsection{Other possibilities}

We have also tested other mechanisms that produce breaks in the
spectral index, but none is able to produce the sharp break seen
in NGC~1569.

Escape from the halo, as described in the last subsection,
in combination with  diffusion as  the dominant propagation mechanism,
produces too weak a
frequency variation of the
radio spectral index.  The reason for this difference
is that  the scale length for
diffusion (i.e. the distance over which CR electrons can
propagate before losing their energy),
$\lambda_{diff}(E)\propto (D(E)/(E b))^{1/2}$,
depends only weakly on energy, even for a constant diffusion coefficient
$D(E)=D_0$. If the diffusion coefficient depends on energy
as $D(E)=D_0 E^\mu$ with $\mu\approx 0.5$, as derived for the
Galaxy \citep{ber90}, the energy dependence of
$\lambda$ is reduced even more.
In the case of convection  the scale length
has a stronger energy dependence, $\lambda_{conv}(E)\propto (V/(E b))$,
and therefore produces a sharper break.

In a closed halo model, a break in the spectrum can be caused
by different energy loss processes that are important for
different ranges of CRE energies.
A break in the synchrotron spectrum of
$\triangle \alpha= 0.5$ can be caused by
adiabatic and bremsstrahlung losses
in addition
to synchrotron and inverse Compton losses.
In some spiral galaxies this can explain the flattening
of the synchrotron spectrum towards lower frequencies
\citep{poh91a}. Taking additionally into account
ionization losses, that are important
at low frequencies,  a total break
of $\triangle \alpha=1.0$ can be produced.
This process is  able to explain
the low-frequency flattening of the spectrum of NGC~4631 \citep{poh91b}.
We have tested these possibilities and 
even in the latter case, the spectrum has much too weak a curvature and is
therefore not able to explain the
sharp bend in the synchrotron spectrum of NGC~1569.

A further possibility to explain a flat low-frequency spectrum are
absorption processes affecting the synchrotron emission. Whereas
synchrotron self-absorption is not important \citep{dee93},
free-free absorption can play a role. We have tested both
absorption by a foreground layer and absorption of material mixed with
the emitting material (see \citealt{dee93}, \citealt{poh91b}) and
found that in neither case
this process is able to explain the sharp break
in the spectrum.

\section{Discussion}

\subsection{Star formation history}

The maps of thermal radio and synchrotron emission trace the
star formation history on different time-scales: Whereas the
thermal radio emission traces the star formation
of the last $\sim 10^7$ years, the synchrotron emission
probes a longer timescale. The lifetime of CREs  in
NGC~1569 range, with the parameters for $B$ and $\urad$ adopted here,
between a few up to about 10 Myr for electrons radiating in the GHz range.
This gives, added to the lifetime of SN progenitor stars of
up to 30 Myr, a span of 40 Myr that is probed by the synchrotron
emission.

The image of the thermal radio emission
in Fig.~\ref{therm-syn-break} shows that the recent SF is concentrated
to a large extent in one peak, coinciding with
the \hii \ region 2. This position furthermore coincides with
the peak of the dust emission (Lisenfeld et al. 2002) and
is about 10 arcsec north of the peak of the CO emission
\citep{tay99} showing that this whole area is active in star formation.

The peak of the synchrotron emission in the south-east shows a region that
has been the site of star formation in the past.
We can derive the SFR from the integrated flux of this region (8 mJy)
and derive,  using eq. 21 in \citet{con92}, $8.3 \times 10^{-4}\, \ms$/yr
for stars more massive than  5 $\ms$.
This is a factor 24 less
than the SFR derived from the peak of the
thermal radio emission (20 mJy and  $2 \times 10^{-2}\, \ms$/yr, using
eq. 23 in \citet{con92}) showing that the SFR in the synchrotron peak
has been less important than the present-day SF in the west of the
galactic disk.
Within the disk, the SF traced by the synchrotron emission has a bar-like
shape and extends in between the SSC A and B.
The ages of these clusters -- the stellar populations are estimated
to have ages up to 9 Myr \citep{gon97} -- are younger than
the SF traced by the synchrotron emission.
This might indicate that the formation of the SSC has been triggered by the
preceding epoch of close-by SF.

The break in the integrated synchrotron emission can  be
explained by the effect of variations of  the SF in the past.
We have shown that it cannot be due the end of the starburst
 5 Myr ago, as suggested by \citet{isr-deb88}. The break could
only be explained by the
beginning of the starburst about 10 Myr ago.
This explanation is however not in agreement with several other
studies that agree that the SFR has declined during the last
5-10 Myr  and that the starburst
preceding this decline had a duration of about 100 Myr
(\citealt{val96}, \citealt{gre98}).
Therefore we conclude that the abrupt beginning of the SF some
10 Myr ago, necessary to explain the break in the
synchrotron spectrum, is unlikely to have taken place.

\subsection{Outflow from the halo}

Since varying SF
has been excluded as a cause of the break in the
synchrotron spectrum, the only  possible explanation is
escape of the CREs from the halo.
Such an outflow is not unexpected:
massive star formation and the related energy input
by SN explosions can lead to an outflow of matter from
the galactic disks into the halo.
In starbursting dwarf galaxies like NGC~1569 this process
is expected to be particularly important due to their relatively small
gravitational potential.
The unusually large size of the radio halo of NGC~1569 is
by itself a sign of a high
energy input by SNe \citep*{dah95}.
Further strong evidence for the presence of a galactic wind
is the weak steepening of the spectral
index in the halo indicating
a convective outflow of CRs along vertical magnetic field lines
which enables a fast (with respect to the energy
loss time scale of the CREs) transport of CRs into the halo.
Such a  magnetic field perpendicular to the disk can be produced by
stretching the field along the flow direction of a galactic wind
\citep{les89}. It has been found in the starburst galaxy
M~82 \citep{reu92}. Recent polarized radio continuum observations
at 8.5 GHz (VLA) and 10.4 GHz (Effelsberg) 
show indeed an outwardly directly magnetic field in  
NGC~1569 \citep{muh03}, supporting the possibility of a convective
outflow from NGC~1569.

Further support for the outflow hypothesis is a general similarity between the
radio continuum map at  1.5 GHz and the \halpha \ image, where the
irregular halo to the south and a long arm extending from the
northwest of the galaxy is clearly visible in both images.
At this low frequency the radio emission is predominantly non-thermal.
An overall similarity  between
the \halpha \ and radio emission in the halos has also been found in NGC~891
\citep*{dah94}, NGC~4631 and NGC~5775 \citep{det92}.
Both emissions are related to regions of active star
formation within the
galactic disk \citep{dah94}. A plausible explanation
for this coincidence is an outflow of ionized gas
due to the high star formation activity which at the same time drags
the magnetic field along and enables therefore a
convective outflow of the CREs
(\citealt{dah95}, 2001).

From our fit to the total synchrotron spectrum
we derived a convection velocity
of 150 \kms. This value is higher than the escape velocity of this galaxy
of 80 -- 110 \kms \citep{mar98}  suggesting that the outflow
is able to eject gas from the galaxy.
An alternative estimate for the convective velocity can be made from the
spectral index maps (Fig. \ref{spectral-index}).
In some filamentary structures
the spectral index does
not steepen outwardly, indicating that the CREs do not suffer
substantial energy losses.
Assuming that the CR electrons were produced
in the disk and taking an average length of such filaments
in the maps of $\alpha_{1.5-5\rm GHz}$ (where the contribution by the
thermal radio emission is still small)
of 20 -- 30$''$ (corresponding
to 220 -- 330 pc) we can derive from the energy loss time-scale
of $4.5 \times 10^6$ yr at 5 GHz
a lower limit for the convective velocity of 50 -- 70 \kms, consistent
with the value derived from the total spectrum.

Observations of the hot gas in the halo of NGC~1569
 provide independent evidence.
\citet{hec95} have found indications for a
supernova-driven outflow from ROSAT X-ray and \halpha \ emission.
They find that about 25\% of the \halpha \ is in a large
($\sim 2$ kpc) complex system with radial velocities of as much
as 200 \kms. This velocity agrees well with our convection velocity
derived from the total spectral index. 
A comparison of the \halpha \ shell speeds to
the escape velocity estimated from the rotation curve and taking into account
a dark matter halo, \citet{mar98} showed that NGC~1569 is likely
to present a substantial loss of shell material from the galaxy.
The velocity of the hot gas traced by X-ray emission is 
well above the escape velocity  \citep{mar99}.
\citet{mar02}
confirmed this conclusion with deep Chandra spectra and found that
the outflow is strongly metal-enriched, carrying nearly all the
metals ejected by the starburst.

\section{Conclusions and summary }

We have present high-sensitivity radio continuum observations at
1.5, 4.9, 8.4 and
15.4 GHz of the dwarf irregular galaxy NGC~1569 at a high resolution
(6 arcsec). These data allow us to perform a spatially-resolved
spectral fitting analysis of the continuum emission from which we
derive maps of the  thermal and synchrotron emission.
The total flux density of the thermal radio emission allows us to
derive the integrated synchrotron spectrum.
The results from  our data and its analysis are:

\begin{itemize}

\item The radio data show an
extended, irregularly shaped halo with filamentary structures around the
galaxy. The spectral indices between 1.5 and 4.9 GHz and between 1.5 and 
8.4 GHz
show an unusually patchy distribution with regions of flat spectrum
extending into the halo.

\item   The distribution of the thermal and synchrotron emission
trace the star formation history on different time-scale
($\sim 10^7$ yr for the thermal emission and $\sim 4\times 10^7$ yr for
the synchrotron emission) in NGC~1569. The thermal
radio emission is mainly concentrated in the brightest \hii region
west of  the super star cluster A and B whereas 
the synchrotron emission  peaks in a
bar-like structure in the disk extending  in between the two star clusters.
The super star clusters are very young ($<10^7$ yr, \citealt{gon97})
 and formed
after the SF period traced by  the synchrotron emission.
It is thus possible that their formation
has been triggered by the older SF in the bar.

\item We confirm the
break in the synchrotron spectrum that was found by \citet{isr-deb88}.
We discuss various possibilities that could produce such a break and
find only  two that are consistent with the radio data: an abrupt beginning
of the star formation about $10^7$ years ago or a convective wind transporting
CREs into an open  halo.
An abrupt end of a starburst several Myr ago, as suggested by
\citet{isr-deb88}, is not able to explain the data because
even in this extreme case the SN rate would not have declined immediately but
only over a period of about $3\times 10^7$ years. This time scale is
longer than the energy loss time scale of CREs so that a sharp
break would be considerably smoothed.
An abrupt beginning of the starburst $10^7$ years ago is not
consistent with data at other wavelengths from which a decline
(and not an increase) in the SFR over this  time scale is derived
(e.g. \citealt{val96}, \citealt{gre98}).
We thus conclude that the only mechanism able to fit the radio data and
being consistent
with data at other wavelengths is a convective wind allowing
CREs to escape from the halo.
This is in agreement with X-ray studies indicating the presence
of a galactic wind (\citealt{hec95}, \citealt{mar02}).

\end{itemize}

\section*{ACKNOWLEDGMENTS}
We would like to thank the referee, U. Klein, for useful suggestions
which helped us to improve the paper.
UL acknowledges support  by the Spanish MCyT  AYA2002-03338 and the
Junta de Andaluc\'\i a (Spain).

{}

\label{lastpage}


\begin{thebibliography}{}
\bibitem[\protect\citeauthoryear{Ables}{1971}]{abl71}
Ables, H. D., 1971, Publ. U.S. Naval Obs. Sec. Ser. XX (IV), 61
\bibitem[\protect\citeauthoryear{Anders et al.}{2003}]{and03}
Anders, P., de Grijs, R., Fritze-v. Alvenslegen, U., Bissantz, N.,
2004, MNRAS, 347, 17 
\bibitem[\protect\citeauthoryear{Arp \& Sandage}{1985}]{arp85}
Arp, H. C., Sandage, A. R., 1985, AJ, 90, 1163
\bibitem[\protect\citeauthoryear{Baron} {1992}]{bar92}
Baron, E., 1992, MNRAS, 255, 267
\bibitem[\protect\citeauthoryear{Berezinsky et al.} {1990}]
{ber90} Berezinsky, V.S., Bulanov, S.V.,
Ginzburg, V.L., Dogiel, V.A., Ptsuskin, V.S., 1990, Astrophysics of
Cosmic Rays, North Holland:Amsterdam
\bibitem[\protect\citeauthoryear{Becker, White \& Edwards}
{Becker et al.} {1991}]{bec91}
Becker, R. H., White, R. L., Edwards, A. L., 1991, ApJS, 75, 1
\bibitem[\protect\citeauthoryear{Clemens, Alexander \& Green}
{Clemens et al.}{1999}]{cle99}
Clemens, M.S., Alexander, P., Green, D.A., 1999, MNRAS, 307, 481
\bibitem[\protect\citeauthoryear{Condon} {1983}]{con83}
Condon, J. J., 1983, ApJS, 53, 459
\bibitem[\protect\citeauthoryear{Condon} {1987}]{con87}
Condon, J. J., 1987, ApJS, 65, 485
\bibitem[\protect\citeauthoryear{Condon et al.} {1998}]{con98}
Condon, J. J., Cotton, W. D.,
 Greisen, E. W., Yin, Q. F., Perley, R. A., Taylor, G. B.,
1998, AJ, 115, 1693
\bibitem[\protect\citeauthoryear{Condon }{1992}]{con92} 
Condon, J. J., 1992, ARAA, 30, 575
\bibitem[\protect\citeauthoryear{Dahlem, Dettmar \& Hummel}{Dahlem et al.} 
{1994}]{dah94}
Dahlem, M., Dettmar, R.-J., Hummel, E., 1994, \AandA, 290, 384
\bibitem[\protect\citeauthoryear{Dahlem, Lisenfeld \& Golla}
{Dahlem et al.} {1995}]{dah95}
Dahlem, M., Lisenfeld, U., Golla, G., 1995, ApJ, 444, 119
\bibitem[\protect\citeauthoryear{Dahlem et al.} {1995}]{dah01}
Dahlem, M., Lazendic, J. S.,
 Haynes, R. F., Ehle, M., Lisenfeld, U., 2001, \AandA, 374, 42
\bibitem[\protect\citeauthoryear{Dettmar} {1992}]{det92} Dettmar, R.J., 1992,
Fund. of Cosmic Phys., 15, 143
\bibitem[\protect\citeauthoryear{Deeg et al.}{1993}]{dee93}
Deeg, H.J., Brinks, E., Duric, N., Klein, U., Skillman, E., 1993,
ApJ, 410, 626
\bibitem[\protect\citeauthoryear{de Vaucouleurs, de Vaucouleurs \& Pence}
{de Vaucouleurs et al.}{1974}]{vau74} de Vaucouleurs, G., 
de Vaucouleurs, A., \& Pence, W., 1974, ApJ, 194, L119
\bibitem[\protect\citeauthoryear{Gonz\'alez Delgado et al.} {1997}]
{gon97} Gonz\'alez Delgado, R.M.,
Leitherer, C., Heckman, T., Cervi\~no, M., 1997, ApJ, 483, 705
\bibitem[\protect\citeauthoryear{Greggio et al.} {1998}]{gre98}
Greggio, L., Tosi, M., Clampin, M.,
de Marchi, G., Leitherer, C., Nota, A., Sirianni, M., 1998, ApJ, 504, 725
\bibitem[\protect\citeauthoryear{Gregory \& Condon }{1991}]{gre91}
Gregory, P. C., Condon, J. J.,
1991, ApJS, 75, 1011
\bibitem[\protect\citeauthoryear{Heckman et al.} {1995}]{hec95}
Heckman, T.M., Dahlem, M., Lehnert,
M.D., Fabbiano, G., Gilmore, D., Waller, W.H., 1995, ApJ, 448, 98
\bibitem[\protect\citeauthoryear{Howarth} {1990}]{how90}
Howarth, N. A., 1990, PhD thesis, University
of Cambridge
\bibitem[\protect\citeauthoryear{Hummel }{1980}]{hum80}
Hummel, E., 1980, \AandA S, 41, 151
\bibitem[\protect\citeauthoryear{Hunter et al.} {2000}]{hun00}
Hunter, D.A., O'Connell, R.W.,
Gallagher, J.S., Smecker-Hane, T.A., 2000, AJ, 120, 2383
\bibitem[\protect\citeauthoryear{Israel} {1988}]{isr88}
Israel, F. P., 1988, \AandA, 194, 24
\bibitem[\protect\citeauthoryear{Israel \& van der Hulst}{1983}]{isr83}
Israel, F. P., van der Hulst, J. M., 1983, AJ, 88, 1736
\bibitem[\protect\citeauthoryear{Israel \& de Bruyn }{1988}]{isr-deb88}
Israel, F. P., de Bruyn, A. G., 1988, \AandA, 198, 109
\bibitem[\protect\citeauthoryear{Israel \& Mahoney} {1990}]{isr-mah90}
Israel, F. P., Mahoney, M.J., 1990, ApJ, 352, 30
\bibitem[\protect\citeauthoryear{Israel \& van Driel} {1990}]{isr90}
Israel, F. P., van Driel, W., 1990, \AandA, 236, 323
\bibitem[\protect\citeauthoryear{Jaffe \& Perola} {1974}]{jaf74}
Jaffe, W.J., Perola, G.C. 1974, \AandA, 26, 423
\bibitem[\protect\citeauthoryear{Kennicutt} {1983}]{ken83} Kennicutt, R.C., 1983, ApJ, 272, 54
\bibitem[\protect\citeauthoryear{Kennicutt} {1984}]{ken84}
Kennicutt, R.C., 1984, ApJ, 277, 361
\bibitem[\protect\citeauthoryear{Klein \& Gr\"ave} {1986}]{kle86}
Klein, U., Gr\"ave, A., 1986,
\AandA, 161, 155
\bibitem[\protect\citeauthoryear{Lesch et al.} {1989}]{les89}
Lesch, H., Crusius, A.,
Schlickeiser, R., Wielebinski, R., 1989, \AandA, 217, 99
\bibitem[\protect\citeauthoryear{Lisenfeld et al.} {1996}]{lis96}
Lisenfeld, U., Alexander, P.,
Pooley, G.G., Wilding, T., 1996, MNRAS, 281, 301
\bibitem[\protect\citeauthoryear{Lisenfeld et al.} {1998}]{lis98}
Lisenfeld, U., Alexander, P.,
Pooley, G.G., Wilding, T., 1998, MNRAS, 300, 30
\bibitem[\protect\citeauthoryear{Lisenfeld et al.} {2002}]{lis02}
Lisenfeld, U., Israel, F. P.,
Stil, J., Sievers, A., 2002, \AandA, 382, 860
\bibitem[\protect\citeauthoryear{Martin}{1998}]{mar98}
Martin, C.L., 1998, ApJ, 506, 222
\bibitem[\protect\citeauthoryear{Martin}{1999}]{mar99}
Martin, C.L., 1999, ApJ, 513, 156
\bibitem[\protect\citeauthoryear{Martin, Kobulnicky \& Heckman}
{Martin et al.} {2002}]{mar02}
Martin, C. L., Kobulnicky, H. A., Heckman, T. M., 2002, ApJ, 574, 663
\bibitem[\protect\citeauthoryear{McCucheon} {1973}]{mcc73}
McCucheon, W.H., 1973, AJ, 78, 18
\bibitem[\protect\citeauthoryear{M\"uhle}{2003}]{muh03}
M\"uhle, S., 2003, PhD thesis, University of Bonn
\bibitem[\protect\citeauthoryear{Niklas et al.} {1995}]{nik95}
Niklas, S., Klein, U., Braine, J.,
 Wielebinski, R., 1995, \AandA S, 114, 21
\bibitem[\protect\citeauthoryear{Pfleiderer, Durst \& Gebler}
{Pfleiderer et al.} {1989}]{pfl80}
Pfleiderer, J., Durst, C., Gebler, K.H., 1980, MNRAS, 192, 635
\bibitem[\protect\citeauthoryear{Pohl, Schlickeiser \& Hummel}
{Pohl et al. }{1991a}]{poh91a}
Pohl, M., Schlickeiser, R., Hummel, E., 1991a, \AandA, 250, 302
\bibitem[\protect\citeauthoryear{Pohl, Schlickeiser \& Lesch}
{Pohl et al.}{1991b}]{poh91b}
Pohl, M., Schlickeiser, R., Lesch, H., 1991b, \AandA, 252, 493
\bibitem[\protect\citeauthoryear{Reuter et al.} {1992}]{reu92}
Reuter, H.-P., Klein, U., Lesch, H., Wielebinski, R., Kronberg, P.P.,
1992, \AandA,250, 302
\bibitem[\protect\citeauthoryear{Stil \& Israel} {1998}]{sti98}
Stil, J., Israel, F.P., 1998, \AandA,
337, 64
\bibitem[\protect\citeauthoryear{Stil \& Israel }{2000}]{sti00}
Stil, J., Israel, F.P., 2000, \AandA,392, 473
\bibitem[\protect\citeauthoryear{Sulentic} {1976}]{sul76}
Sulentic, J. W., 1976, ApJS, 32, 171
\bibitem[\protect\citeauthoryear{Tomita, Ohta \& Saito}{Tomita et al.} 
{1994}]{tom94}
Tomita, A., Ohta,K., Saito, M., 1994, PASJ, 46, 335
\bibitem[\protect\citeauthoryear{Taylor et al.} {1999}]{tay99}
Taylor, C.L., H\"uttemeister, S.,  Klein, U., Greve, A., 1999, \AandA, 349, 424
\bibitem[\protect\citeauthoryear{Vallenari \& Bomans}{1996}]{val96}
Vallenari, A., Bomans, D. J., 1996, \AandA, 313, 713
\bibitem[\protect\citeauthoryear{Waller}{1991}]{wal91}
Waller, W. H., 1991, ApJ, 370, 144
\end{thebibliography}
\end{document}